\documentstyle[11pt,newpasp,twoside,epsf]{article}
\def\edcomment#1{\iffalse\marginpar{\raggedright\sl#1\/}\else\relax\fi}
\reversemarginpar

\begin{document}
\title{Survey of Large Planetary Nebulae in Decay}
\author{Thomas Rauch}
\affil{Institut f\"ur Astronomie und Astrophysik, Univ.\,T\"ubingen, Germany}
\author{Elise Furlan}
\affil{Institut f\"ur Astrophysik, Univ.\,Innsbruck, Austria}
\author{Florian Kerber}
\affil{ST-ECF, Garching, Germany}
\author{Miguel Roth}
\affil{Carnegie Observatories, Pasadena, USA}

\begin{abstract}
The Planetary Nebulae (PNe) return nuclear processed stellar material back to the
interstellar medium (ISM) and thus have an important influence on the chemical
evolution of our Galaxy. We present results of a survey of PNe in decay which
have reached a density comparable to the ambient ISM which leads to an
interaction with it. This gives us the opportunity to investigate properties
of the ISM. We have identified about 20 new examples for this interaction,
demonstrating that it is a more common phenomenon than previously expected:
Different stages of interaction, ranging from the early (asymmetric brightness
distribution) to the very advanced (parabolic or distorted shape and/or an
off-center central star) are obvious.
\end{abstract}

\section{Introduction}
PNe are the result of mass loss of Asymptotic Giant Branch (AGB) stars. They are successfully 
explained in terms of the interacting-stellar-winds model by Kwok, Purton, \& Fitzgerald (1978) as the product 
of the stellar mass-loss and wind history (Dwarkadas \& Balick 1998) during the AGB and post-AGB evolution.

PNe play an important role in the chemical evolution of our galaxy: The amount of nuclear
processed material which is returned to the ISM by supernovae and PNe is about equal during
the first Gyr of galaxy evolution but later the PNe material dominates by about an order of magnitude. 

Large PNe in decay provide a unique tool to study the mixing process in the complex 
PN$\leftrightarrow$ISM interaction zones and, moreover, give us the opportunity to
investigate the properties of the ambient ISM.

\begin{figure}[ht]
\plottwo{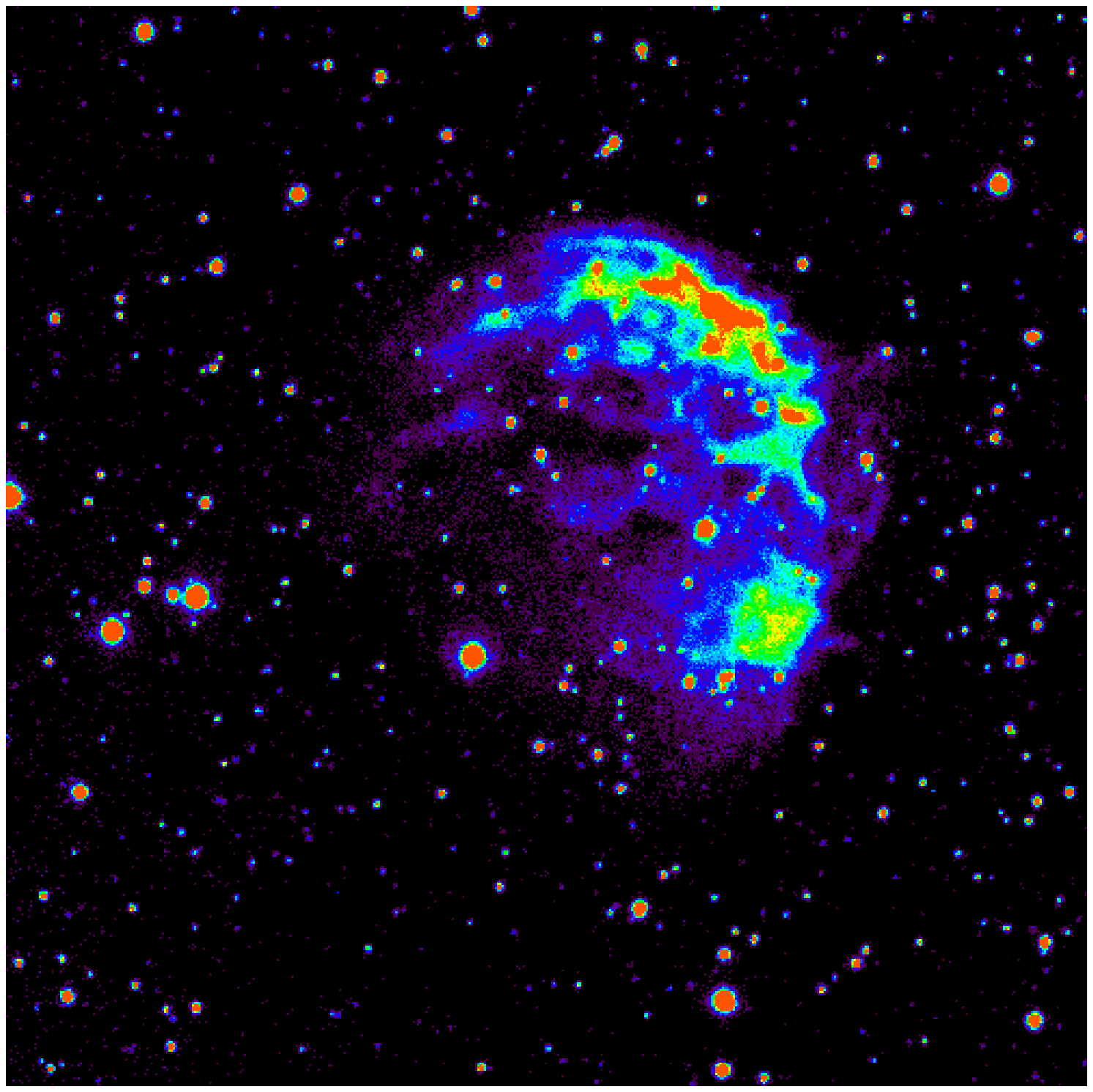}{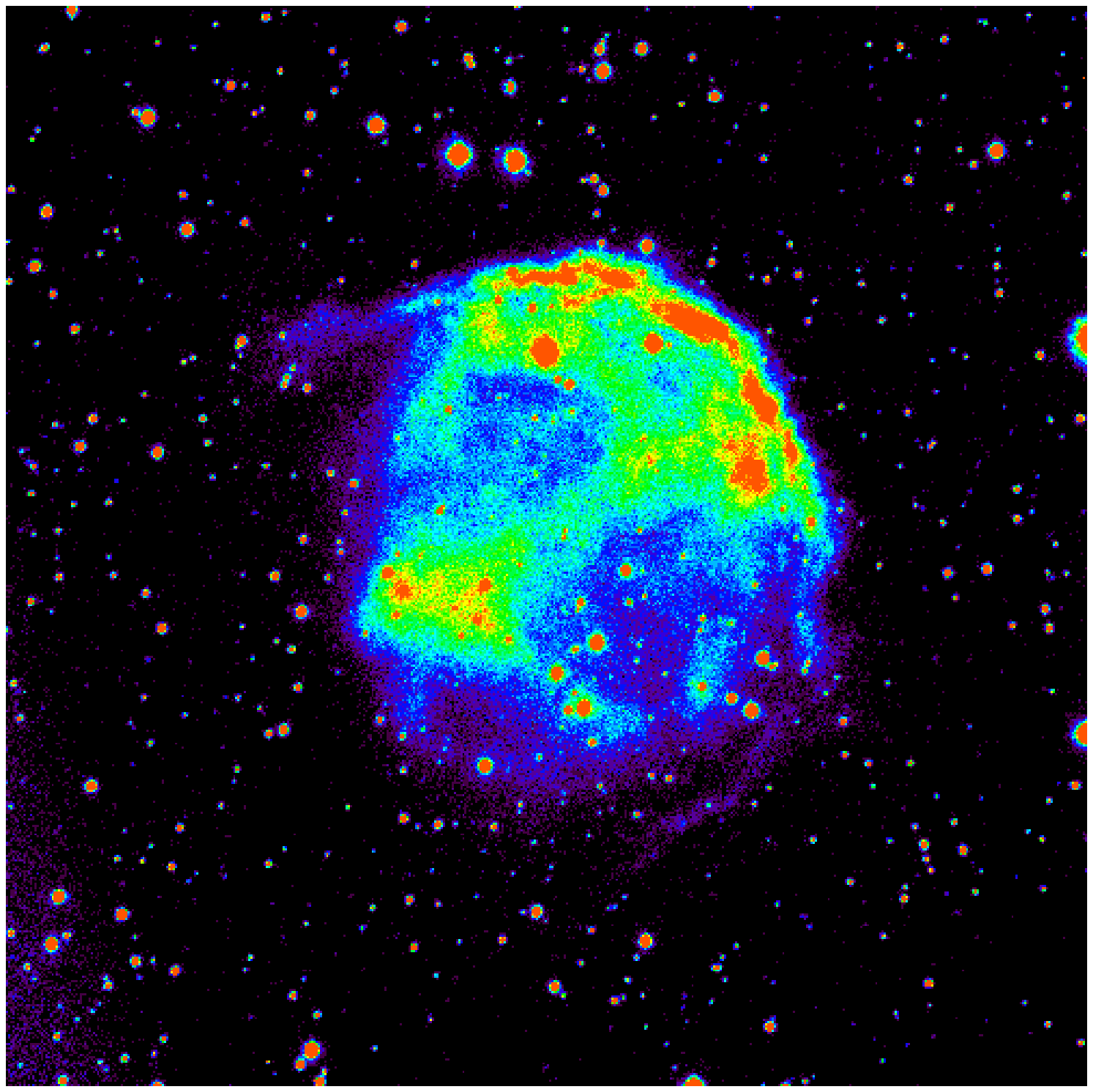}
\caption{
left: KFR\,1 -- The Jelly-Fish nebula. A semi-circular interaction front can be
seen in the northwest with filamentary structures to the southeast.
The CS is de-centered, located very close to the front at the east of the bright radial
structure.
H\,$\alpha$ image (5.5$\arcmin\times$5.5$\arcmin$, exposure time 900\,sec),
north is up and east is left in all images;\\
right: MeWe\,1-1. 
This PN is not as advanced in its decay as KFR\,1. A central bar is prominent as well as two
filamentary tails that prolong from the northwestern interaction front towards the east and
the south.
[N\,{\sc ii}] image (3.8$\arcmin\times$3.8$\arcmin$, 600\,sec)
}
\label{fone}
\end{figure}

\section{Observations}
Over the years interaction of PNe with the ambient ISM has been recognized as a very
important process in the late stage  of PN evolution. Once considered a
curiosity only seen in a few odd examples, like A\,35 (Jacoby 1981; Hollis et al.\,1996)
it has now become evident that this process is much more common.
In our ongoing survey we have collected the largest, homogeneous data set on
old PNe (Kerber et al. these proceedings) interacting with the ISM by means of narrow-band imaging and long-slit
spectroscopy. 

The PN observations (about 25 nights of imaging and ten nights of spectroscopy) have been performed since
1996 at ESO (La Silla, Chile, 2.2m MPG telescope), Las Campanas (LCO, Chile, 1m and 2.5m telescope), DSAZ (Calar Alto,
Spain, 1.23m and 2.2m telescopes), and Wise Observatory (Israel, 1m telescope). 
Central star (CS) spectra were taken at the DSAZ (four nights at the 3.5m telescope), and we were awarded three nights
in Feb.\,2000 at ESO (3.6m telescope) for CS spectroscopy.

Of the 21 objects studied about 75\,\% show signs of
interaction (cf.\,Kerber et\,al.\, these proceedings).
This unexpectedly large percentage may be the result of an
observational bias: the interaction leads to an --- usually asymmetric ---
brightness enhancement in these low surface brightness objects facilitating their
discovery.
With our new survey data we are able to identify the full range of phenomena
associated with the interaction. This makes it possible to distinguish different stages of interaction,
see below. Some examples are shown in Figs.\,1, and 2 (the images shown were
taken at the 100$\arcsec$ ``du Pont'' telescope at LCO in Feb.\,1998).

\begin{figure}[ht]
\plottwo{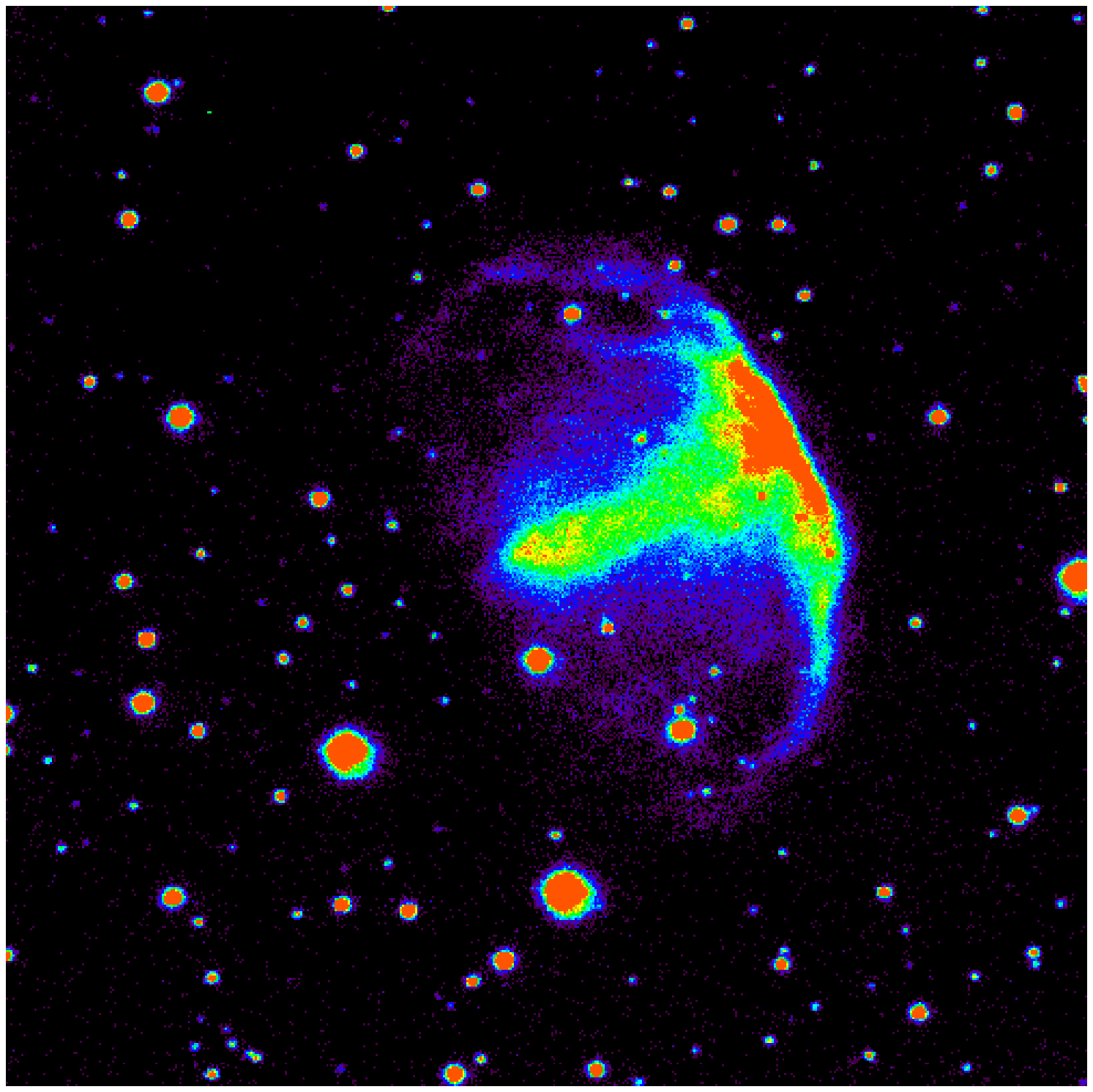}{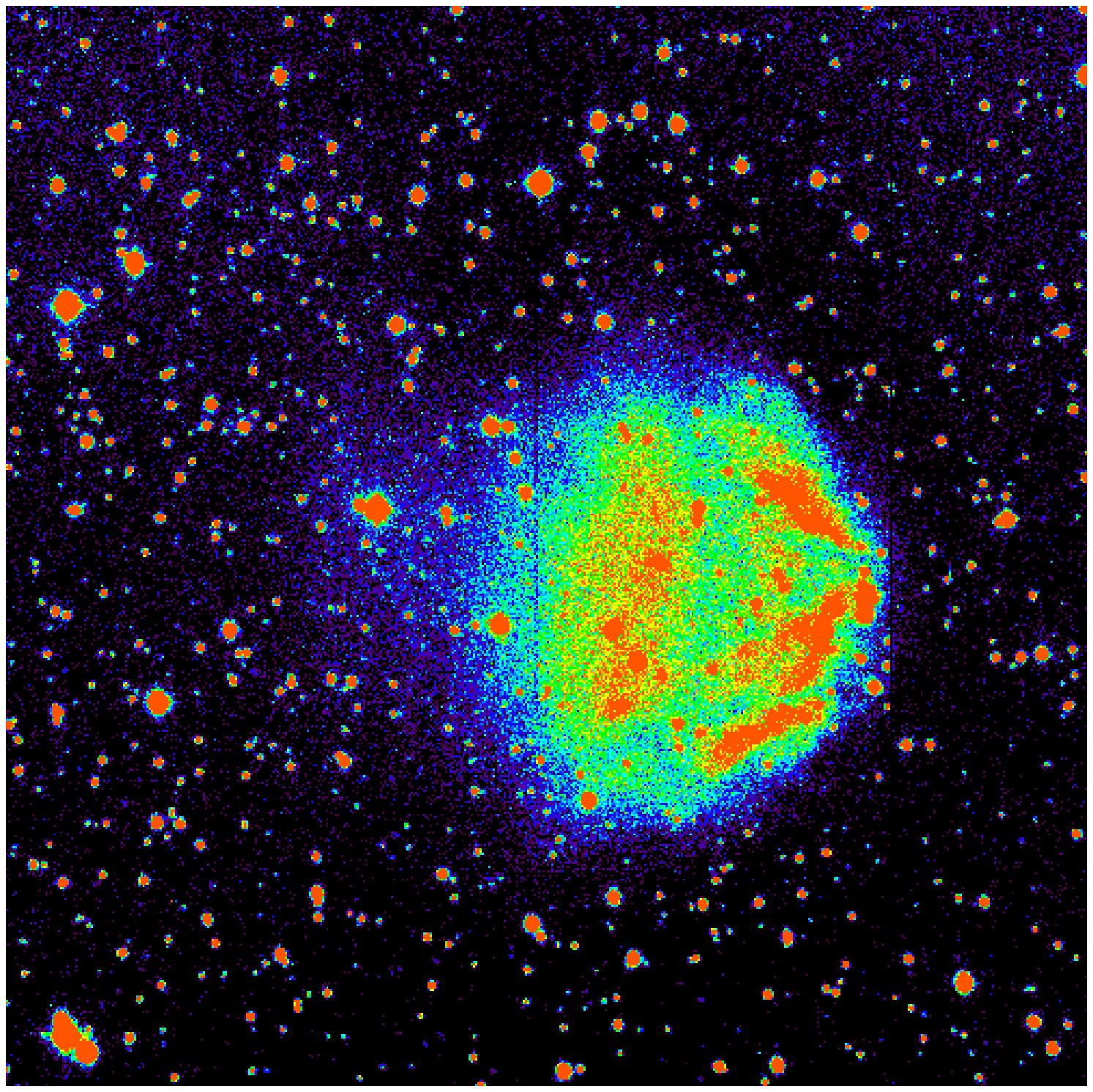}
\caption{
left: SuWt\,1 --- The Hammer Nebula. 
This PN appears similar to MeWe\,1-1. 
SuWt\,1 probably is a former bipolar PN in an
advanced stage of interaction.
H\,$\alpha$ image (1$\arcmin\times$1$\arcmin$, 600\,sec);\\
right: MeWe\,1-4. 
In this roughly circular PN we can see a ``broken'' interaction front
at the southwest due to Rayleigh-Taylor instabilities. The ISM is here
flowing through the PN shell into the inner nebula. Towards the northeast
a long nebular tail is visible due to ``stripping'' off PN matter by the
ISM. The CS is displaced by half of the PN radius towards the western end of the PN.
H\,$\alpha$ + [N\,{\sc ii}] image (3.9$\arcmin\times$3.9$\arcmin$, 1\,800\,sec)
}
\label{ftwo}
\end{figure}

\section{The Interaction Process}
A theoretical basis for the understanding of the interaction process
has been laid by Borkowski, Sarazin, \& Soker (1990)
and Soker, Borkowski, \& Sarazin (1991). When the PNe are moving with respect to the ISM,
we can distinguish different stages of interaction:

\begin{itemize}
\item[$\bullet$]     {\bf young PNe}
\begin{itemize}
\item[$\rightarrow$] density of PN $\gg$ density of ISM
\item[$\rightarrow$] free expansion of PN
\item[$\rightarrow$] no significant influence of ISM
\end{itemize}
\item[$\bullet$]     {\bf mid-age PNe}
\begin{itemize}
\item[$\rightarrow$] density of PN decreases
\item[$\rightarrow$] ISM pressure upstream $\approx$ pressure in PN shell
\item[$\rightarrow$] PN shell is compressed
\item[$\rightarrow$] increase of PN density and surface brightness, asymmetric brightness distribution
\item[$\rightarrow$] higher recombination rate $\rightarrow$ lower degree of ionization
\item[$\rightarrow$] Mach number still large $\rightarrow$ shape of PN not significantly affected, largely spherical
\end{itemize}
\item[$\bullet$]     {\bf old PNe}
\begin{itemize}
\item[$\rightarrow$] PN density drops further
\item[$\rightarrow$] expansion is slowed upstream $\rightarrow$ deformation of PN (blunt parabola) \\ (Fig.\,1: MeWe\,1-1)
\item[$\rightarrow$] Rayleigh-Taylor instabilities lead to severely distorted PN shape \\ (Fig.\,2: MeWe\,1-4)
\item[$\rightarrow$] CS moves out of the PN center \\ (Fig.\,1: KFR\,1, Fig.\,2: MeWe\,1-4)
\end{itemize}
\end{itemize}

\section{Future work}
The survey will be extended to larger objects (this makes use of the Wide Field Imager
at the ESO/MPG 2.2m telescope with a field of view of about 30$\arcmin$). The spectroscopic
characterization will be done for all objects. A spectral analysis of the CS will
be performed by means of NLTE model atmospheres (Rauch et al.\, these proceedings) and
provide information about the CS's evolution and its distance. Together with kinematic studies by
high-resolution PN spectroscopy we will be able to derive absolute positions and motions
of the PNe within our galaxy.
 
\section*{Acknowledgment}
This research is supported by the DLR under grant 50\,OR\,9705\,5 (TR),
and in part by the ``Auslandsabteilung'' of the University Innsbruck (EF).

\end{document}